\documentclass[dvips,11pt,a4paper]{elsarticle}

\usepackage{multibib}
\usepackage[T1]{fontenc}         
\usepackage[latin1]{inputenc}
\usepackage{graphicx}
\usepackage{amsmath}
\usepackage{bm}
\usepackage{tikz}
\usepackage{color}
\usepackage{helvet}
\usepackage{courier}
\usepackage{makeidx}
\usepackage{footmisc}
\usepackage{type1cm}         
\usepackage{textcomp}
\usepackage{subfig}
\usepackage{enumitem,xcolor}    
\usepackage{yhmath}
\usepackage{amssymb}
\usepackage{mathrsfs}
\usepackage{amsbsy}

\usepackage{mathptmx}
\usepackage{color}
\usepackage{enumitem,xcolor}    

\DeclareCaptionLabelSeparator{colon}{. }
\usepackage[labelfont={bf},labelsep={colon}]{caption}

\DeclareMathAlphabet\mathbfcal{OMS}{cmsy}{b}{n}

\begin{document}
\thispagestyle{empty}
\bibliographystyle{plain}

\begin{center}
\textcolor{blue}{ \Large  \bf  On Helmholtz-Hodge decomposition of inertia on a  discrete local frame of reference } \\
\vspace{3.mm}
{\bf Jean-Paul Caltagirone } \\
\vspace{3.mm}
{ \small Bordeaux INP - University of Bordeaux  \\
   I2M Institute, UMR CNRS  n\textsuperscript{o}5295 \\
  16 Avenue Pey-Berland, 33607 Pessac Cedex  \\
\textcolor{blue}{\texttt{ calta@ipb.fr }  } }
\end{center}

\textcolor{blue}{\bf Abstract}
The notion of inertial reference frame is abandoned and I replaced it by a local reference frame on which the fundamental law of mechanics is expressed. The distant interactions of cause and effect are modeled by the propagation of waves from one local reference frame to another. The derivation of the equation of motion on a straight segment serves to express the proper acceleration as the sum of the accelerations imposed on it, in the form of an orthogonal local Helmholtz-Hodge decomposition, in one divergence-free and another curl-free contribution. I wrote the inertia term in the form of a gradient of a scalar potential and a dual curl of a vector potential. The adopted formalism opens the way to a reformulation of the material derivative in terms of potentials and allows me to remove the fictitious forces from  continuum mechanics.

The discrete equation of motion, invariant by rotation at constant angular velocity, is used to conserve the angular momentum per unit of mass, in addition to the conservation of energy per unit of mass and acceleration. All the variables in this equation are expressed only with two fundamental units, those of length and time. 

\normalsize

\vspace{3.mm}
\textcolor{blue}{\bf Keywords}

Discrete Mechanics; Acceleration Conservation Principle; Hodge-Helmholtz Decomposition; Local Frame of Reference

\vspace{-3.mm}
\begin{verbatim}
_______________________________________________________________
\end{verbatim}
\vspace{-2.mm}
This article may be downloaded for personal use only. Any other use requires prior permission of the author and AIP Publishing. 
\vspace{1.mm}

J-P Caltagirone, On Helmholtz-Hodge decomposition of inertia on a  discrete local frame of reference, Phys. Fluids, 32, 083604, https://doi.org/10.1063/5.0015837, 2020.
\vspace{-6.mm}
\begin{verbatim}
______________________________________________________________
\end{verbatim}

\vspace{-5.mm}

\normalsize

\textcolor{blue}{\section{Introduction }}

The homogeneity and isotropy of space, as well as the uniformity of time, define a Galilean reference frame or an inertial reference frame. A context more suited to the fundamental law of dynamics leads to the consideration of only real forces, excluding the forces of inertia. In particular, a rotating reference frame is not inertial.
The question discussed here is different: can the equation of motion remain invariant when we consider only the motion of uniform rotation? In continuum mechanics the Navier-Stokes equation is invariant for a rectilinear translational motion at constant velocity, but it is not for rotation. To describe the mechanical equilibrium it is necessary to add to the first component of the Navier-Stokes equation a fictitious acceleration $\omega^2 \: r \: \mathbf e_r$, where $\omega$ is the angular velocity of constant rotation and $ r $ the distance to the axis. In the rotating frame of reference, space is neither isotropic nor homogeneous. The principle of inertia is therefore not strictly verified.

The equations of the mechanics of the continuous media translate the balance between inertia and the forces which make it possible to maintain the motion of a body in rotation. These forces are fictitious; like centrifugal force, they have no reality except for considering that there are particular links, such as a wire connecting the body and the axis of rotation, a pressure in a fluid or even a fictitious gravity. In other words, the inertia of a rotating body is non-zero, which leads us to formulate the material derivative as a function of the chosen coordinate system. In general, the material derivative is not the vector of the material derivatives of each of the components; the complementary terms are compensated by fictitious forces. 

The resolution of the equation of motion must make it possible to find all the components of the solution, velocity but also pressure, and, in the general case, the scalar and vector potentials. For an incompressible flow for example, if the solution in velocity is known, it is not possible {\it a priori} to extract the pressure from it; only solving the equation does this.

This paper addresses the extension of the invariance of equations of mechanics to uniform rotational motion. Classical mechanics, which describes any vector of space by its components in a global frame of reference, inhibits any chance of achieving this. The proposed model is based on a local reference frame attached to a single segment of finite dimension on which the equation of motion is derived. In this context, any change of inertial frame of reference is no longer possible and the observer is alone on his frame of reference, he cannot interpret distant phenomena. However, interactions at a distance will be described from cause to effect, from one local reference frame to another through the propagation of waves. The law of motion will however be the same in all local reference frames.

In discrete mechanics, the equation of motion is invariant in a rectilinear translation, just like the Navier-Stokes equation in classical mechanics. The last obstacle to overcome corresponds to the model of inertia adopted; it is necessary for the material derivative to be null for a uniform rotational motion. Inertia implicitly involves the curvature of space; for example, the geostationary orbit of a satellite in a circular motion around the earth does not involve the curvature of the circle but that of the earth, equal to twice that of the orbital circle. While inertia in a direction is well represented by that of the law of D. Bernoulli, the fact remains that in general it must be explained to express the curvature in a three-dimensional space.

Invariance by rotation, according to Noether's theorem \cite{Noe11}, allows us to conclude that the angular momentum is preserved. It is therefore essential that in addition to the conservation of momentum associated with the translation invariance, the equation of motion also allows to keep the angular mo-mentum. Energy conservation is ensured by the invariance of the time translation. For the phenomena described by the equation of motion to be independent of the frame of reference, it is necessary for all these invariances to be intrinsically ensured by this equation.

The concept of the Galilean frame of reference and that of inertia are well established in the context of modern mechanics and have not attracted much attention for many decades. However, their deep meanings are still, and rightly, discussed by science historians \cite{Rou06, Mar06, Mar07}. The concepts developed here would allow us to revisit certain physical phenomena such as turbulence, where rotation and incessant changes in the trajectory of the fluid play decisive roles in energy transfers between vortices. The particular case of turbulence in rotating flows is very prominent in the literature and applications \cite{God15, Buz18, Bro20, Gal20}; the influence of the formulation of inertial terms could be studied on cases already treated.

By abandoning the classical concept of global inertial reference frame, discrete mechanics makes it possible to interpret inertia as the mean curvature of the inertial potential. It excludes the presence of fictitious forces or accelerations in the equation of motion. In addition, the celerity of waves (tides, acoustic, electromagnetic) is a local property, which makes it possible to state that the laws of physics are the same in all local frames of reference regardless of the celerity of the medium (particle, material medium or vacuum).

\textcolor{blue}{\section{\label{dismech:level1}Discrete mechanics framework }}

One of the main differences between continuum mechanics and discrete mechanics is the abandonment of the notion of continuous medium and global inertial frame of reference in favor of a local frame of reference.  
This non-standard notion is clarified in the context of discrete mechanics. Both velocity $\mathbf V =\mathbf V^o + \bm \gamma \: dt$ and displacement $\mathbf U = \mathbf U^o + \mathbf V \: dt$ are relative quantities, the functions of acceleration $\bm \gamma$ and elapsed time $dt$. Knowing their current values requires knowledge of their values $\mathbf V^o$ and $\mathbf U^o$ at an earlier time $t^o$. 
Acceleration $\bm \gamma$ is considered, on the other hand, as an absolute, local and measurable quantity on an oriented segment at all times. The localization of a particle or a material medium in space is no longer necessary, as information circulates from one point to another in a discrete neighborhood of cause and effect; the discrete horizon is limited to a distance equal to $dl = c \: dt$ where $c$ is the celerity of the wave (sound or light). The concepts of absolute and Galilean inertial frame of reference become useless, since all the interactions are located in the immediate vicinity of an oriented segment on which accelerations, velocities and displacements are expressed.
The derivation of the equation of motion is carried out on this concept of local frame of reference; this filters and immediately eliminates all uniform movements of translation and rotation, re-establishing the concept of Galilean invariance locally and extending it to rotational motions.
The independent quantities $dl$ and $dt$ are chosen according to the problem posed, they must be adapted to the physics to be understood. The spatial scale can be reduced as much as necessary but the geometric structure cannot in any event be reduced to a point in order to preserve the direction of the segment. The discrete motion equation is supposed to represent the physical phenomena from field theory at all spatial and temporal scales.

\textcolor{blue}{\subsection{\label{vectcal:level2}Acceleration in discrete vector calculus }}

\begin{figure}[!ht]
\begin{center}
\includegraphics[width=6.cm]{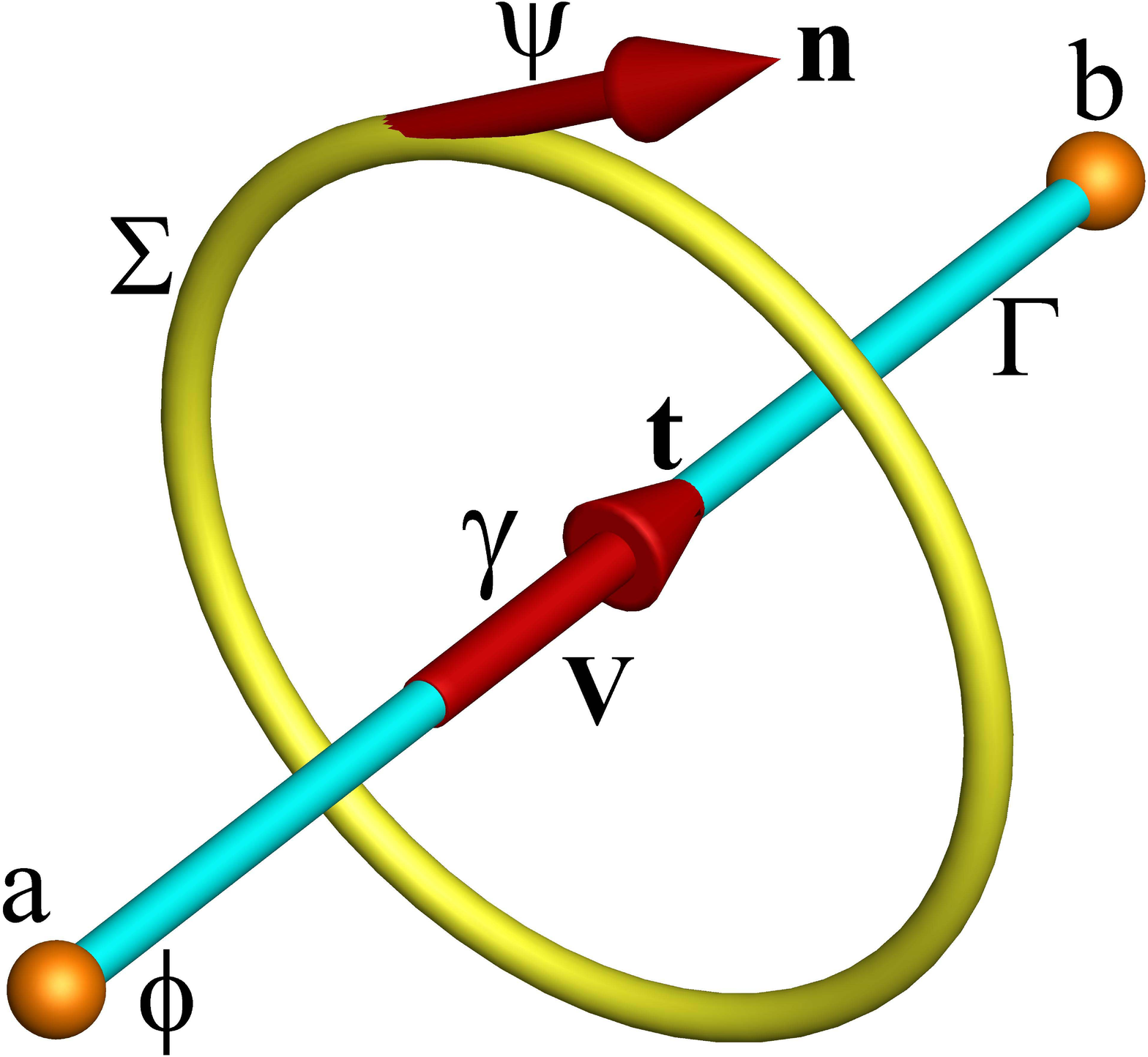}
\caption{\it Elementary pattern of discrete geometric topology: the own acceleration on the segment $\Gamma$ is due to two contributions, (i) the direct acceleration due to a difference of the inertial potential $\phi$ between $a$ and $b$, (ii) the acceleration induced by the circulation of the vector potential $\bm \psi$ along the contour $\Sigma$.  }
\label{primdual-a}
\end{center}
\end{figure}

Let us consider the geometric topology represented in figure (\ref{primdual-a}), made up of a rectilinear segment $\Gamma$ limited by the two points $a$ and $b$; this segment is oriented according to the unit vector $\mathbf t$. The dual geometric topology corresponds to an oriented contour $\Sigma$ along the vector $\mathbf n$ such that $\mathbf n \cdot \mathbf t = 0$. In general, the scalar potentials $\phi$ will be attached to the points $a$ and $b$ and the vector potentials $\bm \psi$ defined on the contour $\Sigma $. The polar vectors, or rather their components, will be associated with the oriented segment $\Gamma$, like acceleration $\bm \gamma$ and velocity $\mathbf V$, where they are assumed to be constant. These vectors will only exist on this primal geometrical topology, while the axial vectors will be defined on the dual contour.

The objective is to define the notion of inertia in this discrete framework and to research how to describe the variation of velocity as a function of time in the frame of reference attached to the segment.
In a Lagrangian description of motion, as in continuum mechanics, the partial derivative of velocity with respect to time is equal to the material derivative, $\partial \mathbf V / \partial t = d \mathbf V / dt$. In a Eulerian description where the reference frame is fixed, the partial derivative can no longer be obtained from the classical expressions of continuum mechanics, in particular those based on the gradient of velocity, since this notion no longer exists.

It is important to note that the quantities $\bm \gamma$ and $\mathbf V$ are components of vectors projected on an oriented segment $\Gamma$. While $\nabla \phi$ and $\nabla \times \bm \psi$ are carried by the same segment, they are not the same components as those of the corresponding vectors. This difference, with the representation of the variables resulting from the notion of continuous medium on global coordinates, is essential. Mechanical equilibrium is no longer defined by an equality on each of the components of the equation of motion.

The equation of motion will be derived over a segment and the vectors as such will not be known explicitly but defined only by their components. 

In the framework of continuum mechanics, the inertial term coming from the material derivative is written $\mathbf V \cdot \nabla \mathbf V$ or $\nabla \cdot \left (\mathbf V \otimes \mathbf V \right) - \mathbf V \: \nabla \cdot \mathbf V$ or $\nabla (\| \mathbf V \|^2/2) - \mathbf V \times \nabla \times \mathbf V$. The last term, the vector of Lamb \cite{Lam93}, \cite{Ham08}, is also not adapted to a discrete description of the mechanics \cite{Cal19a} since it is not a dual curl. It is therefore necessary to characterize the inertia in a different way and to rewrite the expression of the material derivative starting from potentials $\phi_i$, the inertial scalar potential and $\bm \psi_i$ the inertial vector potential in the form:
\begin{eqnarray}
\displaystyle{\frac{\partial \mathbf V}{\partial t} = \frac{d \mathbf V}{ d t} - \nabla \phi_i + \nabla \times \bm \psi_i }
\label{temporal}
\end{eqnarray}

Physically the variation of velocity over time on the segment $\Gamma$ with respect to the material derivative i.e. following the segment during its motion, is due to two actions, the direct one associated with the scalar potentials at both extremities and the other induced by the circulation of the vector potential along the contour $\Sigma$. The material derivative is fixed by the accelerations imposed from the outside and thus the velocity variation is reduced by a quantity representing the inertia. The proper acceleration of the material medium or the particle then takes the form:
\begin{eqnarray}
\displaystyle{ \bm \gamma = \frac{d \mathbf V }{dt}  \equiv \frac{\partial \mathbf V}{\partial t} + \nabla \phi_i - \nabla \times \bm \psi_i =  \frac{\partial \mathbf V}{\partial t} + \bm \kappa_i}
\label{material}
\end{eqnarray}
where $\bm \kappa_i$ is called inertia, a constant polar vector on the segment $\Gamma$. For a steady motion, the acceleration is reduced to inertia alone.

The choice of the inertial potential $\phi_i$ naturally tends towards its definition given by Bernoulli where, in his law, the variation of pressure along a streamline is equal to that of $\| \mathbf V \|^2/2$. In the present framework of discrete mechanics, the mass is an abandoned quantity and the pressure replaced by the potential $p / \rho$ where $\rho$ is the local density. The vector potential is not independent of $\phi_i$, it is quite simply $\bm \psi_i = \phi_i \: \mathbf n = \| \mathbf V \|^2/2 \: \mathbf n$. The inertia induced by $\bm \psi_i$ on the segment $\Gamma$ is due to the action of the other segments of the global primal topology constituting the physical domain.

The material derivative (\ref{material}) is not expressed by component as in continuum mechanics; it is not possible to transform inertia from one form to another with a vector calculus formula.
The inertia vector $\bm \kappa_i$ carried by $\Gamma$ can be interpreted as the curvature of the potential $\phi_i$, defined as a potential surface. The quantity $\kappa_i = \kappa_l + \kappa_t$ defined at a point on this surface is the mean curvature calculated on two orthogonal unit vectors, $\mathbf t$ and $\mathbf t \times \mathbf n$. The mean curvature is by definition independent of the orientation of these. If we name $f (x, y, z)$ the implicit function corresponding to the potential surface and $H$ the Hessian matrix of the second derivatives, the curvature of the surface of normal $\mathbf n$ at a point in a direction $\mathbf t$ is equal to $\kappa = Hes(\mathbf n, \mathbf t) / \| \mathbf n \| $, where $Hes (\mathbf n, \mathbf t) = \mathbf t^t \: H \: \mathbf t$ is the Hessian. The two contributions of $\kappa_i$ are the longitudinal curvatures $\kappa_l$ and transverse $\kappa_t$.

The vector $\bm g = \nabla \phi_i$ is transformed into $\bm r = \nabla \times \bm \psi_i$ using a rotation matrix $R = ([0, -1], [1,0])$ corresponding to an angle of $\pi / 2$ around the normal $\mathbf n$ at each facet $\mathcal S$. These vectors are orthogonal $\bm g \cdot \bm r = 0$ and the norm of their external product is equal to $\| \bm g \wedge \bm r \| = \| \bm g \| \: \| \bm r \|$ according to the identity of Lagrange, i.e. $\| \bm g \wedge \bm r \|^2 = (\| \bm g \| \: \| \bm r \|)^2 - \| \mathbf g \cdot \mathbf r \|^2$.
The inertial acceleration induced on the segment $\Gamma$ is the result of all the contributions of each of the facets $\mathcal S$ and the component of the inertial acceleration $\bm \kappa_i$ on the support $\Gamma $ is the sum of direct and induced accelerations.

Thus the physical notion of inertia presented here differs from those of continuum mechanics. The tensor concept  initially developed to unify, in particular, solid and fluid mechanics, but also used later for the theory of relativity, served to find a general formal framework for the theory of fields \cite{Lan71b}. Despite this successful generalization effort, it turns out that the formulations obtained cover and mix the two direct and induced fundamental actions, as in the Navier-Stokes equations. 
Formulations with potentials like those of the Maxwell and Navier-Lam{\'e} equations clearly dissociate the longitudinal effects linked to $\nabla \cdot \mathbf V$ from the transverse effects associated with $\nabla \times \mathbf V$.
 The Helmholtz-Hodge decomposition of acceleration into a curl-free component and a divergence-free component is the cornerstone of discrete mechanics.
\begin{figure}[!ht]
\begin{center}
\includegraphics[width=6.cm]{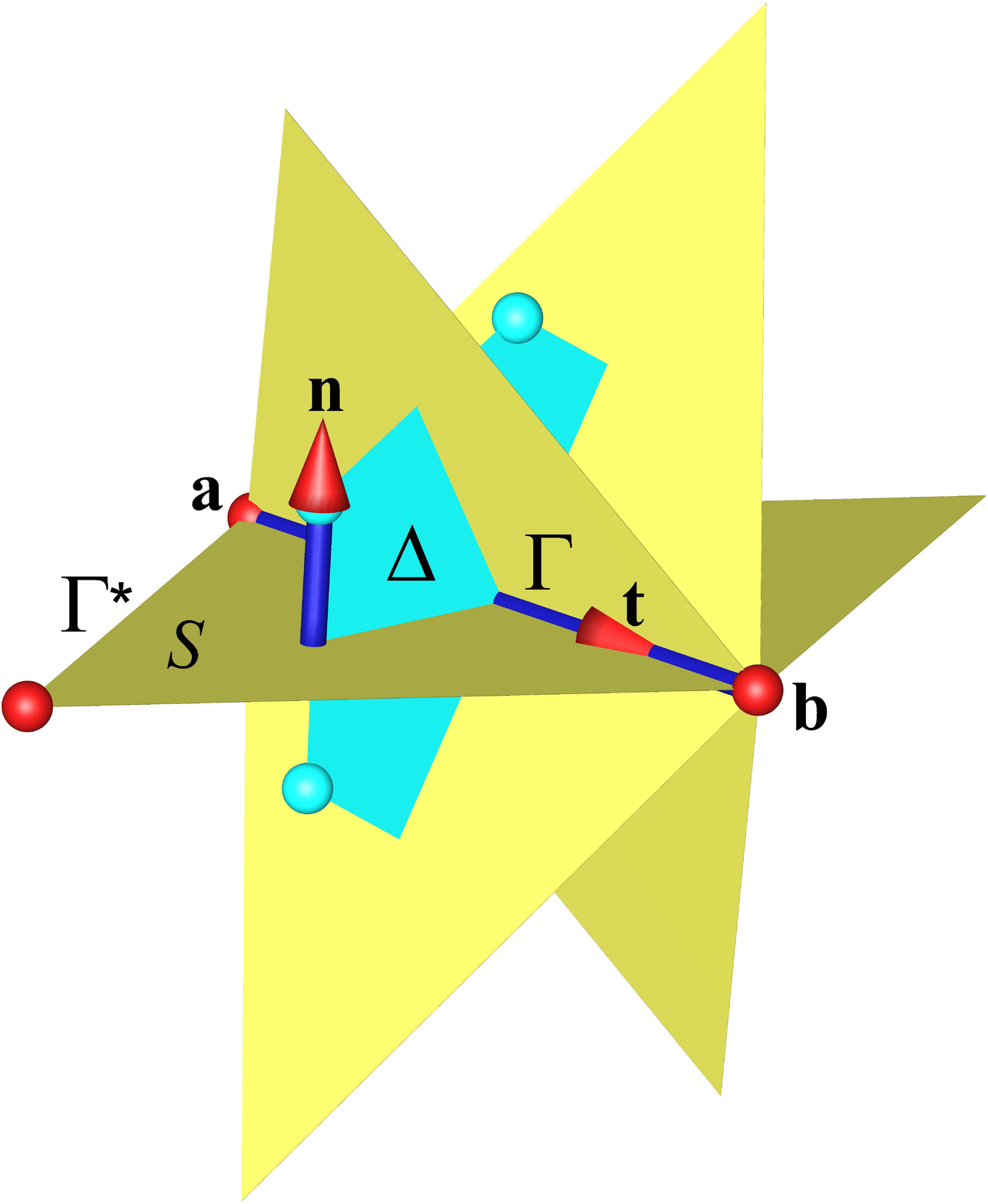}
\caption{\it Stencil of geometric topology formed by six facets $\mathcal S$ around the common segment $\Gamma$; the dual planar surface $\Delta$ is orthogonal to this segment.  }
\label{primdual-b}
\end{center}
\end{figure}

The construction of the geometric topology is continued by the connection of segments between each other by the vertices to obtain, for example, a primal facet $\mathcal S$ starting from the contour $\Gamma^*$ represented in figure (\ref{primdual-b}). The dual contour $\Sigma$ also makes it possible to define the flat surface $\Delta$ used to calculate the fluxes of the quantities associated with $\Gamma$. The stencil formed by several facets having the segment $\Gamma$ in common shows its central role; in a larger structure composed of a set of stencils, each segment will participate symmetrically, in the direct effects for itself, and in the induced effects for the others.

We show that the global structure obtained has important properties \cite{Cal20a}: (i) the vectors $\nabla \phi$ and $\nabla \times \bm \psi$ are locally orthogonal, (ii) the discrete operators mimic the properties of the continuous, $\nabla_h \times \nabla_h \phi = 0$ and $\nabla_h \cdot \nabla_h \times \bm \psi = 0$ regardless of the regular functions $\phi$ and $\bm \psi$.

\textcolor{blue}{\subsection{\label{noniner:level2} Non-inertial flows}}

For certain flows the inertia terms are strictly zero; this is the case for Couette or Poiseuille flows, for example. In some cases the terms are both equal to zero and in others each of them is non-zero but their sum is indeed equal to zero.

Let us consider, for example, the velocity field $\mathbf V = (1 - y^2) \: \mathbf e_x$ corresponding to the flow of a Newtonian fluid between two undefined planes subjected to a constant pressure gradient.
In continuum mechanics the Poiseuille flow is not inertial, but the reasons differ according to the form of the chosen term:
\vspace{2.mm}
\begin{itemize}
\item $\mathbf V \cdot \nabla \mathbf V$: the inertia is identically zero because $\mathbf V$ and $\nabla \mathbf V$ are two orthogonal vectors;
\vspace{-2.mm}
\item $\pmb{\mathscr{L}} = - \mathbf V \times \nabla \times \mathbf V$: this Lamb vector is not zero but it is equal and opposite to $\nabla \left (\| \mathbf V \|^2/2 \right)$; the sum is zero on each of the components.
\end{itemize}

The form of inertia contained in the proposed material derivative (\ref{material}) leads to two non-zero terms, but in discrete mechanics, the equilibrium is not described by an equality on each component; it is maintained locally by the constraints imposed, here the incompressibility of the Poiseuille flow. In conclusion, there is no inertia for this motion, but for different reasons from those mentioned in continuum mechanics.

\textcolor{blue}{\subsection{\label{dismot:level2} Discrete motion equation}}

The equation of motion is derived on the segment $\Gamma$ from Galileo's intuitions, the principle of equivalence of gravitational effects and inertia, and the principle of Galilean relativity where velocity is only defined to a close translational motion. In discrete mechanics \cite{Cal19a}, \cite{Cal20a}, mass is excluded from the equation of motion; if $\bm h$ represents the set of external accelerations projected on the segment, then the proper acceleration will be written in the form $\bm \gamma = \bm h$, an equality between accelerations. The Helmholtz-Hodge decomposition \cite{Des05}, \cite{Gue06}, \cite{Bha13}, \cite{Ang13}, applies to this as to any other vector, so it follows that the equation of motion becomes:
\begin{eqnarray}
\displaystyle{ \bm \gamma = - \nabla \phi + \nabla \times \bm \psi }
\label{dechh}
\end{eqnarray}

The absence of mass or density in the equation of motion poses no theoretical difficulty. The equivalence between mass and energy makes it possible to define energies per unit of mass, represented here by the scalar and vector potentials. If each of the terms of the equation of motion is an acceleration, the variable chosen to express them is the velocity or, rather, its components on the segments $\Gamma$:
\begin{eqnarray}
\left\{
\begin{array}{llllll}
\displaystyle{ \bm \gamma = - \nabla \left( \phi^o - c_l^2 \: dt \: \nabla \cdot \mathbf V \right) + \nabla \times \left( \bm \psi^o - c_t^2 \: dt \: \nabla \times \mathbf V \right) +  \bm h_s  } \\  \\
\displaystyle{ \alpha_l \: \phi^o - c_l^2 \: dt \:  \nabla \cdot \mathbf V  \longmapsto \phi^o  } \\ \\
\displaystyle{\alpha_t \: \bm \psi^o - c_t^2 \: dt \: \nabla \times \mathbf V \longmapsto \bm \psi^o } \\ \\
\displaystyle{   \mathbf V^o + \bm \gamma \: dt \longmapsto \mathbf V^o  }
\end{array}
\right.
\label{discrete}
\end{eqnarray}
where $c_l$ and $c_t$ are the longitudinal and transverse celerities of the media concerned, fluid, solid or vacuum; the velocity $\mathbf V$ is a function of the acceleration and the velocity at the previous instant, $\mathbf V = \mathbf V^o + \bm \gamma \: dt$, and the displacement is written $\mathbf U = \mathbf U^o + \mathbf V \: dt$. The symbol $\longmapsto$ corresponds to an explicit upgrade of the quantity concerned after solving the vector equation of the system (\ref{discrete}). The factors $\alpha_l$ and $\alpha_t$ between $0$ and $1$ correspond respectively to the attenuations of the longitudinal and transverse waves. The source term $\bm h_s$, for example gravitational, capillary effects, etc., will be written in the same way, in the form of a Helmholtz-Hodge decomposition. All the variables and the physical parameters of this system are expressed by the two fundamental units, length and time.

The system of equations (\ref{discrete}) constitutes a model with continuous memory which stores energy information between $0$ and $t$. The energies per unit of mass of compression $\phi^o$ and of shear-rotation $\bm \psi^o$ are upgraded at each stage in time and can be described by the integrals:
\begin{eqnarray}
\displaystyle{   \phi^o = - \int_0^t \: c_l^2 \: \nabla \cdot \mathbf V \:  d\tau;  \:\:\:\:\:\:\:\:\:\:
 \bm \psi^o = - \int_0^t \:  c_t^2  \: \nabla \times \mathbf V \: d\tau } 
\label{disint}
\end{eqnarray}

The initial instant must correspond to a state of mechanical equilibrium, i.e. the vector equation of the system (\ref{discrete}) must be verified identically. In practice, it is possible to choose the relative rest state where the motions are associated with a translation or a rotation of rigid body. The own acceleration is the material derivative $d \mathbf V / dt$ and the displacement $\mathbf U = \: \mathbf V \: dt$ which serves to express the vector equation in terms of velocity and thus to imply all the terms. The nonlinear inertia terms contained in the material derivative will themselves be linearized.

The autonomous nature of this formulation should be emphasized here: the solution is obtained without any constitutive law or additional conservation law. In particular, the conservation of the mass is not necessary to obtain a solution, as is the case for the Navier-Stokes equation to which it is added. This formulation in $(\bm \gamma, \phi^o, \bm \psi^o)$ allows us to evaluate {\it a posteriori} the velocity, the displacement, etc. It presents itself as a law of conservation of energy per unit of mass, where $\phi^o$ is the compression energy and $\bm \psi^o$ the shear-rotation energy. The energy-mass duality could possibly favor this latter but it would unnecessarily increase the number of fundamental quantities.

It is possible to consider an incompressible and irrotational flow whose acceleration of gravity is written $\mathbf g = - \nabla \phi_g + \nabla \times \bm \psi_g$. We thus obtain a generalization of Bernoulli's law:
\begin{eqnarray}
\displaystyle{  - \nabla \left( \phi^o + \frac{\| \mathbf V \|^2}{2} + \phi_g \right) + \nabla \times \left( \bm \psi^o + \frac{\| \mathbf V \|^2}{2} \: \mathbf n + \bm \psi_g \right) = 0  }
\label{bernoul}
\end{eqnarray}

The scalar potential replaces the pressure $\phi^o = p / \rho$ and the vector potential translates the acceleration of rotation. It should be noted that, in the terminology of continuum mechanics, a vector derives from a potential when only the scalar potential is considered. Gravitational acceleration is derived from two potentials, like all other vectors. All source terms then take this form. In the case of a particular streamline, the first term of the relation (\ref{bernoul}) is used to find the law of Bernoulli and, as the operator's gradient and dual curl are orthogonal, the constant is the same for the two terms.

For compressible irrotational flows, only the first term of the equation (\ref{discrete}) is preserved; note that the celerity of the sound $c_l$ can depend on different variables. This equation of motion is not associated with any constitutive law, it is intrinsic; only the celerities and the relaxation factors must be known.

\textcolor{blue}{\section{\label{invar:level1} Invariances}}

\textcolor{blue}{\subsection{\label{galinv:level2} Galilean invariance for an uniform rotational motion}}

The definitions of a Galilean frame of reference cover different notions depending on whether they are associated with invariance, transformations or equations. They differ from the principle of material frame indifference stated by Truesdell $\&$ Noll \cite{Tru92}, \cite{Spe98} which is considered as a constraint for the constitutive laws in continuum mechanics. Controversies over this concept still persist when it is extended to balance sheet equations.

The subject here relates to the rotation motion at constant angular velocity which, unlike the translation motion at uniform velocity, does not satisfy the original principle of inertia; indeed, it generates fictitious forces to ensure its balance. In fact, there is sometimes a confusion between the state and the process, i.e. the state is that which corresponds to a solely kinematic vision and the process is the path which leads to this state.

\paragraph{Steady state equilibrium}

Consider a uniform rigid rotation motion with constant angular velocity $\bm \Omega$ imposed from the start around an axis $z$. The stationary angular velocity $V_{\theta}$ takes the form $V_{\theta} = \bm \Omega \times \mathbf r \cdot \mathbf e_{\theta}$ and incompressibility is already ensured, $\mathbf V = 0$ as a hypothesis. From a kinematic point of view alone, it is not possible to invoke a force as a pressure gradient or source term to ensure this constraint. In continuum mechanics, the material derivative is not zero, it is equal to $- \omega^2 \: r \: \mathbf e_r$ where $\omega = \bm \Omega \cdot \mathbf e_z$.
Any radial outward motion due to this contribution leads to an increase in volume and therefore to a non-zero divergence, contrary to the hypothesis adopted.

To balance the uniform rotational motion for an elastic solid it is then necessary to introduce a fictitious acceleration, the centrifugal acceleration equal to $\bm \Omega \times \bm \Omega \times \mathbf r$ on the same component. Similarly, in a fluid medium, we could define a pressure $p = \rho \: \omega^2 \: r^2/2$.
These cases go beyond the hypothesis of the kinematic point of view. In classical mechanics, mechanical balance is only ensured by the action of fictitious forces.
Thus, in discrete mechanics the material derivative is zero $d \mathbf V / dt = 0$, intrinsically and locally.

\paragraph{Path to mechanical equilibrium}

The equations of discrete mechanics (\ref{discrete}) symmetrically define potentials, the energies of compression and rotation, which are associated with operators $\nabla \cdot \mathbf V$ and $\nabla \times \mathbf V$ which are never perceived as gauge conditions. For example, the condition $\nabla \cdot \mathbf V = 0$ is not imposed, it is obtained as a function of the celerity of the wave, $c_l$ the celerity of the acoustic waves.

Let us re-examine the rigid rotation motion from the point of view of discrete mechanics as a continuous memory problem whose initial condition corresponds to the state of rest. The fluid is supposed to be Newtonian, its celerity $c_l$ is known and its viscosity $\nu$ is constant. During the acceleration motion in rotation, the medium cannot be considered as incompressible and $\nabla \cdot \mathbf V \ne 0$, while the motion of rigid body is indeed divergence-free. This is one of the essential characteristics of discrete mechanics; the motion becomes incompressible at large time constants, it is not {\it a priori}.

The evolution over time of the rotational motion of an elastic solid or a viscous fluid can be considered from the same angle, by calculating the velocity field from a rest state. When the motion is perfectly established there will no longer be any difference in the motion of the rigid body of the two media.

Let us consider a cylindrical tank of axis $Oz$ containing water whose viscosity $\nu = 10^{-6} \: m^2 s ^{-1}$. Initially immobile, the water is rotated by the rotation of the surface of the tank with radius $R = 1 \: m$ at the velocity $\mathbf V \cdot \mathbf e_{\theta} = \omega \: r = V_0$. Obviously the height of the tank following $ Oz $ will be large enough so that the Eckman layers have no influence on the motion of the fluid; it will be driven exclusively subject to viscous radial effects.
The solution to this problem can be easily obtained by solving the equation of motion in the approximation of an incompressible motion. The method of separating the variables leads to a general solution on the difference $v_{\theta}(r, t) = \mathbf V \cdot \mathbf e_{\theta} - \omega \: r$ obtained by superposition, which is written: 
\begin{eqnarray}
 \displaystyle{ v_{\theta}(r,t) = \sum_{n=0}^{\infty} \left( a_n J_1 \left(\alpha_n \frac{r}{R} \right) + b_n Y_1 \left(\alpha_n \frac{r}{R} \right) \right) e^{-\nu \: \frac{\alpha_n^2}{R^2} \: t} }
\end{eqnarray}
where the coefficients $\alpha_n$ are obtained from the boundary conditions using the orthogonality properties of the Bessel functions $J$ and $Y$.
What remains is to satisfy the initial condition, namely that the fluid is immobile for $t = 0$:
\begin{eqnarray}
\displaystyle{  - \frac{V_0 \: r}{R} =  \sum_{n=0}^{\infty} \left( a_n J_1 \left(\alpha_n \frac{r}{R} \right) + b_n Y_1 \left(\alpha_n \frac{r}{R} \right) \right) }
\end{eqnarray}

The coefficients $a_n$ and $b_n$ are thus determined in turn.
The solution $\mathbf V \cdot \mathbf e_{\theta}(r, t)$ shows the development of the dynamic boundary layer in the vicinity of the solid wall and its thickening over time. 
We can admit that the solution is very precisely described by the first term in Bessel's series development, and limit the term $\alpha_n / R^2$ by unity. We are led to:
\begin{eqnarray}
\displaystyle{ \varepsilon = \mathbf V \cdot \mathbf e_{\theta} - V_0 \frac{r}{R}  \approx e^{-\nu \: t  } }
\end{eqnarray}

If the difference is, for example, $\varepsilon = 10^{-4}$, as the viscosity $\nu = 10^{-6}$ we find an establishment time of $t \approx 10^7 \: s$. In order for the observer to consider that he is dealing with a solid whose characteristic value of the shear modulus is twelve orders of magnitude greater than the viscosity of a fluid, the time constants must be in a ratio reverse. While the solid sets in almost instantaneous rotation when the tank is driven in rotation, the fluid is driven by viscosity over long time constants.
Basically, there is no difference between the behaviors of a Newtonian fluid and an elastic solid, only the observation time $dt$ is different.

Let us now examine the operation of the discrete equation (\ref{rotat}) during a simulation corresponding to the rotation of the fluid:
\begin{eqnarray}
\displaystyle{\frac{\partial \mathbf V}{\partial t} + \nabla \phi_i - \nabla \times \left( \phi_i \: \mathbf n\right) = - \nabla \left( \phi^o - c_l^2 \: dt \: \nabla \cdot \mathbf V \right) + \nabla \times \left(\bm \psi^o - \nu \: \nabla \times \mathbf V \right) } 
\label{rotat}
\end{eqnarray}

During the unsteady resolution of the entrainment of a fluid in a circle of radius unit by the external surface at the velocity $\mathbf V_{\theta} = 1$, the terms of the equation describe different phenomena:

\begin{itemize}
\item the term $\nabla \times \left (\nu \: \nabla \times \mathbf V \right)$ corresponds to the transfer of momentum from outside to inside the circle to give the expected steady solution at convergence. It does not generate any acceleration in the radial direction. As the Newtonian fluid does not accumulate the shear energy $(\alpha_t = 0)$, the instantaneous shear-rotation energy per unit of mass is simply equal to $\bm \psi^o = \nu \: \nabla \times \mathbf V$; at convergence it is equal to $\bm \psi^o = 2 \: \nu$ and its dual curl is zero;
\item the term $- \nabla \phi^o$ represents the compression energy per unit of mass accumulated without dissipation $(\alpha_l = 1)$ throughout the process. When the stationary solution is obtained, the scalar potential is equal to $\phi^o = \omega^2 \: r^2/2$ and the acceleration to $- \nabla \phi^o = - \omega^2 \: r \: \mathbf e_r$;
\item inertial acceleration $\nabla (\| \mathbf V \|^2/2)$ increases from zero to its equilibrium value $\omega^2 \: r \: \mathbf e_r$;
\item the second inertia term $- \nabla \times (\| \mathbf V \|^2/2 \: \mathbf n)$ tends toward the quantity $- \omega^2 \: r \: \mathbf e_{\theta}$ and translates the orthoradial acceleration on a closed contour of radius $r$. As the constraint $dt \: c_l^2 \: \nabla \cdot \mathbf V$ tends to ensure incompressibility, this is transferred to the radial component and therefore represents the centripetal acceleration. This quantity, calculated from a balance between the radial and orthoradial components of the acceleration, is equal to $- 2 \: \omega^2 \: r \: \mathbf e_r$, i.e. the value corresponding to the Lamb radial vector. But in the context described, the inertia $\bm \kappa_i = \nabla \phi_i - \nabla \times \bm \psi_i$ cannot be represented by component.
\end{itemize}

Thus the equation of motion (\ref{discrete}) is a set of operators, each of which represents a well-defined physical phenomenon. The divergence of the velocity $\nabla \cdot \mathbf V$ and its primal curl $\nabla \times \mathbf V$ ensures the link between the velocity and the different accelerations of the equation. Its functioning, more complex than that of classical mechanical equations, reflects the strong interactions between the phenomena. The role played by the potentials makes this equation a continuous memory model.

In continuum mechanics with the Navier-Stokes equation, the equilibrium is satisfied on each of the components. The traditional use of the Lamb vector, $\pmb{\mathscr{L}} = - \mathbf V \times \nabla \times \mathbf V$, in classical mechanics is due to this choice of expressing the laws of mechanics by component in a global reference frame. Unfortunately this vector is not divergence-free and leads to other difficulties, such as that of the meaning of the divergence of this vector \cite{Cal20a}.

In discrete mechanics, the centrifugal radial acceleration is compensated by the orthoradial acceleration, the result of which is centripetal. The constraint $dt \: c_l^2 \: \nabla \cdot \mathbf V$ conserves the mass and acts as an internal acceleration to maintain an average potential (pressure) zero in the domain without boundary conditions. Mechanical balance is inherently ensured without any added fictitious force or additional constitutive law.

\paragraph{Composition of two motions}

The change of reference frame or the composition of motions is accompanied by additional terms due to the non-linearities in the motion equation. Let us examine what becomes the proper acceleration if we consider two movements such as $\mathbf V = \mathbf R + \mathbf W$; suppose that one of the velocities represents a rotational motion at constant angular velocity, $\mathbf R = \bm \Omega \times \mathbf r$.

The right-hand side of the equation remains unchanged, but, taking into account the expression $ \| \mathbf V \|^2 = \| \mathbf R \|^2 + \| \mathbf W \|^2 + 2 \: \mathbf R \cdot \mathbf W$, the material derivative becomes:
\begin{eqnarray}
\displaystyle{\frac{d \mathbf V}{d t} \equiv \frac{\partial  \mathbf V}{\partial t} + \nabla \left( \frac{ \| \mathbf W \|^2}{2} \right) -  \nabla \times \left( \frac{\| \mathbf W \|^2}{2} \: \mathbf n \right)  + \nabla \left(  \mathbf R \cdot \mathbf W \right) - \nabla \times \left(  \mathbf R \cdot \mathbf W \: \mathbf n \right) }
\label{compos}
\end{eqnarray}

The two terms in $\| \mathbf R \|^2/2$ integrated within the operators of the right-hand side classically represent the centrifugal acceleration $\bm \Omega \times \bm \Omega \times \mathbf r$.
The last two terms of (\ref{compos}) replace the Coriolis acceleration $2 \: \bm \Omega \times \mathbf W$ of continuum mechanics, which is not a pure curl. As for the Lamb vector, the application of the divergence or dual rotational operator to the Coriolis vector produces additional terms \cite{Cal20a}. In discrete mechanics, the exterior product is in turn transformed into two terms of a Helmholtz-Hodge decomposition.

\paragraph{Conclusions on rotational motion}

It is therefore important to dissociate the inertial accelerations included in the material derivative from the accelerations imposed by the exterior so as to exclude any reference to fictitious forces to ensure the mechanical balance fixed by the equation of motion. The inertial terms of discrete mechanics make it possible to obtain a local invariance of the translational motions at constant velocity and of the uniform rotational motion. The equation of motion (\ref{discrete}) also shows that the right-hand side is invariant for these two motions.

Beyond the concepts contained in the concepts of the Galilean frame of reference, group of Galilean invariance or material frame indifference, it is the law of motion which must intrinsically have local invariance for these two motions and for their combinations.

\textcolor{blue}{\subsection{\label{invcons:level2} Invariances and conservations laws}}

Noether's theorem \cite {Noe11} allows us to conclude that the invariance of physical systems with respect to spatial translation gives the law of conservation of linear momentum (spacetime symmetries). Likewise, the notion of isotropic space and invariance by rotation stated by E. Noether reflects the fact that the laws of physics do not change when we consider a rigid rotation motion which gives the law of conservation of angular momentum. According to the same theorem, the invariance with respect to time translation gives the law of conservation of energy.

The equation of motion (\ref{discrete}) translates the evolution in time of the mechanical equilibrium from time $t^o$ to $t^o + dt$. The elapsed time is independent of velocity and flows regularly. The compression and rotation energies (\ref{disint}) are intrinsically preserved by the time upgrades from $\nabla \cdot \mathbf V$ and $\nabla \times \mathbf V$.

The conservation of mass is implicitly ensured by the equation of discrete motion. The density is upgraded like the scalar potential by the divergence of the velocity:
\begin{eqnarray}
\displaystyle{ \frac{d \rho}{dt} + \rho \: \nabla \cdot \mathbf V = 0}
\label{masse}
\end{eqnarray}

The mass or the density do not enter explicitly into the equation (\ref{discrete}), which represents at the same time a law of conservation of acceleration and of energy.

\textcolor{blue}{\section{\label{conclusions:level1} Conclusions}}

Inertia plays a very important role in all fields of physics, and in particular for turbulent flows where the spatial and temporal scales of structures are very small. It is therefore essential to be able to characterize the form of the terms which represent inertia within the equation of motion.
The choice of representing inertia as a Helmholtz-Hodge orthogonal decomposition into a curl-free component and another divergence-free component makes it possible to attribute specific properties. For example, the divergence of the dual curl of the inertial potential is identically zero, while the divergence of the Lamb vector is not. These differences lead to interpretations which may be debatable.

The essential result of the formulation proposed on inertia is the invariance of all the terms of the equation of discrete motion with respect to a motion of uniform translation, but also of a motion of uniform rotation. In other words, the Galilean invariance is extended to all motions of translation and rotation and to their linear combinations. This was made possible thanks to the replacement of the concept of global inertial reference frame by a local frame of reference concretized by an segment on which the direct and induced accelerations are projected.

From the practical point of view, this continuous memory formulation requires knowledge of an earlier state of mechanical equilibrium to predict the existence of another later state of equilibrium. The observation of the state of a physical system at a given time is not enough to characterize it, it depends on the path followed.

\clearpage

\vspace{4.mm}
{\bf Author Contributions}

Author: Physical modeling, Conceptualization, Methodology, Research code,  Validation, Writing- Original draft preparation, Reviewing and Editing.

The paper has been checked by a proofreader of English origin.

\vspace{4.mm}

{\bf Declaration of Competing Interest}

There are no conflict of interest in this work.

\vspace{4.mm}

{\bf Acknowledgments}

As professor emeritus, the author thanks the Institute de M{\'e}canique et d'Ing{\'e}ni{\'e}rie de Bordeaux.

\vspace{5.mm}

\nocite{*}
\bibliography{database}

\end{document}